# Designing Memory Bits with Dissipation lower than the Landauer's Bound

**Saurav Talukdar, Shreyas Bhaban, James Melbourne and Murti V. Salapaka***

200 Union Street SE, University of Minnesota, Minneapolis, MN 55455, USA

E-mail: *murtis@umn.edu

March 2018

**Abstract.** A Brownian particle in a symmetric double well potential is used as a representation for a single bit memory, where, the location of the particle in either well denotes one of the two states of a single bit memory. This article analyzes the effect of modifications to a symmetric double well potential on the minimum heat dissipation associated with erasure of the information stored in a single bit memory. Two types of modifications are considered, viz., overlap between the two wells and the asymmetry between the two wells of a bit of memory. Moreover, the analysis presented here, takes into account the uncertainty in the success of the erasure process. We quantify the effect of the proposed modifications on the heat dissipation accompanying erasure of a bit of information with a comparison to the Landauer's bound. In particular, we conclude that the proposed modifications could result in the minimum heat dissipation being lower than the Landauer's bound in quasi-static erasure processes. Furthermore, we quantify the physical aspect ratio for designing a memory bit which could improve memory density.

## 1. Introduction: file preparation and submission

Over the last four decades, the semiconductor industry has made significant headway in improving the performance of CMOS-FET devices, while consistently reducing their size. It has enabled tremendous improvements in the capabilities of personal computers, laptops and smart phones; to the point where a modern smart phone is several hundred times faster than the computational abilities of devices that guided the NASA's travel to the moon in 1969. A crucial enabler is increase in energy efficiency of computing devices without compromising on performance.

Recent research is focused toward investigating fundamental limits on how efficient a device can be or how small its physical dimensions can get; an idea resonating with the limit on heat engine efficiency proposed by Carnot [1]. Subsequent fundamental contributions [2, 3, 4] demonstrated a link between the fields of Thermodynamics and Information Theory, leading to an important understanding of the limits on the consumption of energy in performing bit level operations.

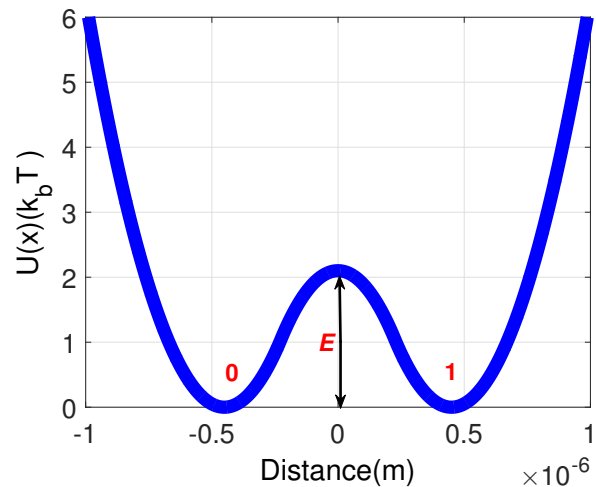

**Figure 1.** A symmetric double well potential, $U(x)$, where the location of the particle in the left and right well designate the state zero and one respectively of a single bit memory. Here, $E$ denotes the barrier height.

A popular model of a single bit memory is a Brownian particle in a symmetric double well potential in a heat bath of constant temperature $T$, with the two identical wells separated by a barrier of sufficient height as shown in Fig. 1. The presence of the particle in either well denotes one of the possible two states of a single bit memory. In this article, we designate the particle's presence in the left well and right well as state zero and one of the memory respectively. The presence of an energy barrier of sufficient height ($\gg k_B T$, $k_B$ is the Boltzmann constant), ensures that information is retained for a long duration, as is desired from a memory bit. Erasure of a bit of information is a computational step, where the outcome is zero irrespective of the initial state of the memory bit. In simple terms, irrespective of whether the particle resides in the right or left well, it needs to be transported to the left well to accomplish a successful erasure. Landauer's principle [2], asserts that, erasure of a bit of information, provided both states of a single bit memory are identical, is accompanied by an average heat dissipation of at least $k_B T \ln 2$. Landauer argued that erasure of information lowers the entropy of the overall system and thus, is to be accompanied by heat dissipation to the surrounding. Bennett further utilized Landauer's argument to explain Maxwell's demon to avoid a paradox violating the second law of thermodynamics [4, 5]. Following the original work of Landauer, there are numerous analytical[6, 7] and experimental studies[8, 9, 10, 11, 12, 13, 14] focused on the minimal energy consumption related to information processing.

A recent interest is on devising erasure mechanisms which result in heat dissipation lower than $k_B T \ln 2$ (also referred to as the Landauer's bound). In this regard, researchers have studied the effect of uncertainty in the erasure process and have shown that partially successful erasures result in heat dissipation lower than the Landauer's bound [15, 16]. It is shown in [15, 16] that an average heat dissipation of $k_B T(\ln 2 + p \ln p + (1-p) \ln(1-p))$, referred as the Generalized Landauer Bound, is associated with erasure of a bit of information stored in a symmetric one bit memory, with $p$ being success proportion of the erasure process. An interesting observation is that, a slight compromise on accuracy (by about 10%) of the erasure lowers the associated minimum heat dissipation significantly (by about 50%). Here, the assumption as made in the previous studies [6, 8, 9, 10, 11, 12, 13, 14] is that, there is 'insignificant' overlap between the two physical states that realize the single bit memory, that is, the two states of a memory bit have sufficient 'physical separation'.

In this article, we analyze the Generalized Landauer Bound (GLB) and its relation to the physical separation between the two states of a one bit memory (that is, size of the memory bit) in the cases where the two states are identical or not identical. We refer to allowing overlap and asymmetry between the two states as introducing 'modifications' in a memory bit. This article analyzes the effect of these modifications on the Generalized Landauer Bound and provides directions for designing memory devices with lower dissipation. Our approach introduces an 'overlap parameter' and uses properties of mixture of Gaussian distributions to derive bounds on change in entropy associated with partial information erasure. The derived bounds converge to the GLB when the overlap between the two states become 'insignificant' and hence is consistent with existing results in the literature [15, 16, 6]. We quantify the relationship between the overlap and the reduction in entropy change and utilize it to arrive at GLB. Here, we derive sub-Gaussian upper bounds analytically as opposed to the numerical study of entropy approximations for bit erasure in [17]. Furthermore, we obtain complimentary lower bounds on the decrease in thermodynamic entropy, demonstrating that these bounds are reasonably sharp; and for bi-stable wells, physically separated by lengths close to their standard deviation, the error in entropy approximation incurred by the 'insignificant overlap' approximation is significant. These quantitative results are of immediate application as they allow a tight approximation of the change in entropy in erasure process, relevant to the precise estimation of the associated minimal heat dissipation. A quantitative analysis is also provided for the case when the two states of memory are non identical and the erasure process moves the state into the well with higher volume. Moreover, we also quantify a physical aspect ratio for designing a memory bit, where a value higher than the one specified in this article is inconsequential from the energetics standpoint and could possibly improve memory density.

## 2. Single Bit Memory and Erasure Process

Following the original work of [2], we consider a Brownian particle in a double well potential as a model for a single bit memory. The location of the particle in either well designates the state of the memory as 0 or 1. In this article, if the particle is located in the left well, we denote the state of memory as 0 and if the particle is located in the right well, we denote the state as 1. The particle in either well is in thermal equilibrium with the surrounding whose temperature is assumed to be a constant $T$. For most part of this article, we assume that both the wells are identical unless specified.

In recent experimental validations of Landauer's principle [8, 18, 9, 12], it is seen that the double well potential in the neighborhood of the two stable states can locally be approximated as a convex quadratic function. It then follows from the Canonical distribution expression that the equilibrium probability distribution is approximately a Gaussian distribution. Henceforth, we assume that the equilibrium probability distributions of the particle in the left and right well are $\mathcal{N}(-\mu, \sigma^2)$ and $\mathcal{N}(\mu, \sigma^2)$ respectively, where $\mathcal{N}(\mu, \sigma^2)$ denotes a Normal distribution with mean $\mu$ and variance $\sigma^2$. Thus, if the memory is in the state 0 the equilibrium probability distribution of the particle is $f_0(x) = Ce^{-(x+\mu)^2/2\sigma^2}$ and if the memory state is 1 the equilibrium probability distribution of the particle state is $f_1(x) = Ce^{-(x-\mu)^2/2\sigma^2}$, where $x$ denotes the position of the particle and $C$ is the normalization constant. This assumption is consistent with the notion of memory as described in [19], where a single bit memory is defined as a system with two stable states, such that, the system is locally in equilibrium in either state and stays in one of the stable states (information retention) for the duration the memory bit is valid (related to the Kramer's escape time across a barrier), after which, stored information is said to be lost.

In this article we study the effect of overlap between the two wells of the double well potential (that is, the effect of overlap between the two equilibrium probability distributions of the two states of a single bit memory) on the Landauer's bound. The two distributions intersect at $x = 0$, where, $f_0(0) = f_1(0) = Ce^{-\alpha^2/2}$ with $\alpha := \mu/\sigma$. Thus, the overlap between the two distributions is characterized by the parameter $\alpha$, which is referred to as the overlap parameter. Higher(lower) the value of $\alpha$, lower(greater) is the overlap between the two equilibrium distributions. Moreover, $\alpha$ is also indicative of the aspect ratio of the memory bit as well as physical extent of the memory bit for a known standard deviation of the equilibrium probability distributions.

Prior to erasure, it is equally likely for the state of the memory, $M$, to be zero or one, that is, $P(M = 0) = P(M = 1) = \frac{1}{2}$. The probability of finding the Brownian particle between $x$ and $x + dx$ is given by,

$$\begin{aligned}
P(X \in (x, x + dx)) &= P(M = 0)P(X \in (x, x + dx)|M = 0) \\
&\quad + P(M = 1)P(X \in (x, x + dx)|M = 1) \\
&= \frac{1}{2}f_0(x)dx + \frac{1}{2}f_1(x)dx.
\end{aligned} \quad (1)$$

Thus, the probability distribution function, $f(x)$, of the particle prior to undergoing an erasure process, is an equally weighted mixture of $f_0(x)$ and $f_1(x)$. In order to represent a valid single bit memory, this distribution must be a bi-modal distribution (due to a symmetric double well stable potential). In this regard, an interesting result from the Statistics community is worth mentioning: an equally weighted mixture of symmetric Gaussian distributions is uni-modal if and only if $\alpha \leq 1$; otherwise it is bi-modal [20]. Thus, in order to represent a symmetric single bit memory with two well defined states, $f(x)$ needs to be a bi-modal distribution with the two modes being alike. Hence, a single bit memory must have $\alpha > 1$ in order to be a valid memory bit.

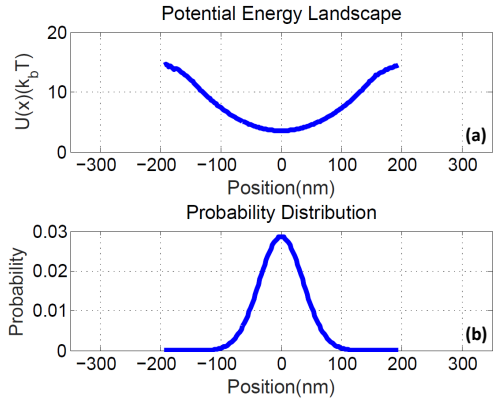

**Figure 2.** (a) Probability distribution and (b) the resulting single well potential obtained from simulations. Here, $\mu = 30\ nm$, $\sigma = 36\ nm$ and thus $\alpha < 1$.

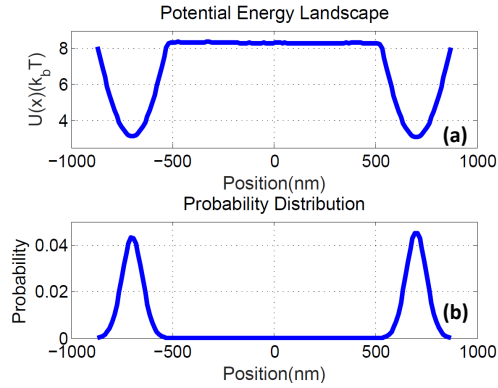

**Figure 3.** (a) Probability distribution and (b) resulting double well potential obtained from simulations. Here, $\mu = 700\ nm,$, $\sigma = 36\ nm$ and thus $\alpha > 1$.

Fig. 2 shows the equivalent probability distribution $\tilde{f}(x)$, obtained by binning 10000 samples drawn at random from $f(x)$, and potential $U(x) = -\ln\frac{\tilde{f}(x)}{C}k_BT$ obtained, when $\mu = 30\ nm$, $\sigma = 36\ nm$(that is, $\alpha < 1$). It is seen that the distribution is uni-modal (the corresponding potential is a single well potential), and cannot be used to realize the two states of a single bit memory. However, $\mu = 700\ nm$ and $\sigma = 36\ nm$ results in $\alpha > 1$, thereby ensuring a symmetric bimodal distribution and a symmetric double well potential as seen in Fig. 3, which results in a well defined memory bit. It is important to note that the overlap parameter $\alpha$ depends on the physical design of the memory bit (the double well potential precisely). The $\alpha > 1$ condition is satisfied in recent studies on Landauer's bound [8, 18, 9, 12].

Erasure is a process where irrespective of the initial state of the memory bit, the final state is zero (also known as reset to zero). With regards to the particle in a double well potential representation of a single bit memory, irrespective of the initial location of the particle, the particle needs to be transferred to the left well. The Generalized Landauer Bound (GLB) from [15, 16, 8] states that, if the probability of success of the erasure process is $p$, then the associated average heat dissipation is at least $k_BT(\ln 2 + p\ln p + (1-p)\ln(1-p))$. The GLB evaluates to $k_BT\ln 2$ for $p = 1$; to 0 for $p = 0.5$ and is shown in Fig. 4 as a function of the success probability $p$. Note that $p < 0.5$ is not considered here as it would imply a reset to one operation. It is important to point out that the success proportion, $p$, depends on the protocol used for accomplishing the erasure. We would like to bring the attention of the reader to [8, 21, 12], where, the authors discuss the dependence of success proportion on protocol parameters.

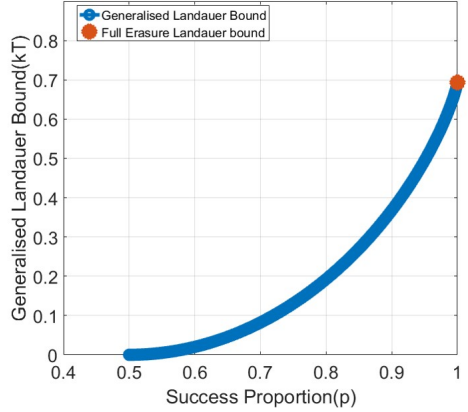

**Figure 4.** GLB as a function of success proportion $p$.

Under the assumption of $\alpha > 1$, we analyze the effect of overlap parameter $\alpha$ on the generalized Landauer bound. The probability distribution function of the particle before undergoing erasure, $f(x)$, is given by, $f(x) = \frac{1}{2}f_0(x) + \frac{1}{2}f_1(x)$ (see eq. (1)). After applying an erasure protocol to the memory bit with success probability $p$, the probability of finding a particle between $x$ and $x + dx$ is given by,

$$P(X \in (x, x+dx)) = pf_0(x)dx + (1-p)f_1(x)dx. \qquad (2)$$

Let, $g(x) := pf_0(x) + (1-p)f_1(x)$. The thermodynamic entropy of the system before erasure is, $S_f = -k_B \int_{-\infty}^{\infty} f(x) \ln(f(x)) dx$, and after undergoing erasure process with a success proportion $p$ is, $S_g = -k_B \int_{-\infty}^{\infty} g(x) \ln(g(x)) dx$. It follows from the $2^{nd}$ Law of Thermodynamics that the average heat dissipation,

$$\langle Q_d \rangle \geq T(S_f - S_g) = k_B T(I_1 - I_2),$$

where, $I_1 = \alpha^2 - \frac{1}{\sqrt{2\pi}\alpha} e^{-\frac{\alpha^2}{2}} \int_{-\infty}^{\infty} e^{-\frac{x^2}{2\alpha^2}} \cosh(x) \ln(\cosh(x)) dx$ and $I_2 = \alpha^2 - \frac{1}{\sqrt{2\pi}\alpha} e^{-\frac{\alpha^2}{2}} \int_{-\infty}^{\infty} e^{-\frac{x^2}{2\alpha^2}} (pe^{-x} + (1-p)e^x) \ln(pe^{-x} + (1-p)e^x) dx$.

The dependence of $I_1 - I_2$ on the overlap parameter $\alpha$, is shown in Fig. 5. It is seen that $k_B T(I_1 - I_2)$ approaches the GLB for large values of the overlap parameter $\alpha$. Note that for $\alpha \leq 2.1$, the decrease in thermodynamic entropy, $k_B(I_1 - I_2)$ is less than $k_B(p \ln p + (1-p) \ln(1-p) + \ln 2)$ and is approximately equal to $k_B(p \ln p + (1-p) \ln(1-p) + \ln 2)$, when $\alpha > 5$. Thus, as seen from Fig. 5, for a given success proportion $p$ and $1 < \alpha < 5$, the associated minimum average heat dissipation due to erasure is lower than the GLB. Hence, introducing overlap (by reducing the physical separation) between the two states of the memory bit can enable energy dissipation lower than the GLB in erasure processes.

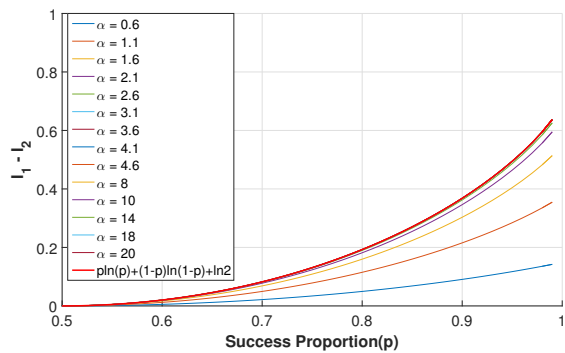

**Figure 5.** $I_1 - I_2$ as a function of $p$ for various values of $\alpha$. $I_1 - I_2$ approaches $p \ln p + (1-p) \ln(1-p) + \ln 2$ (red curve) as $\alpha$ increases.

We would like to mention that the success proportion $p$ is dependent on the protocol used for performing erasure. The reader should refer to [8, 12] to understand that success proportion is dependent on a protocol parameter and the nature of the double well potential. So, in order to achieve the same success proportion with different values of $\alpha$, the protocol needs to be changed accordingly. Designing of erasure protocols which maintains the proportion of success for various values of the overlap parameter is beyond the scope of this article.

We now derive upper and lower bounds to the change in entropy as a function of the success proportion $p$ and separation parameter $\alpha$, that converge to the GLB as the parameter $\alpha \to \infty$.

Consider, $Y_g \sim g(.) := pN(-\mu, \sigma^2) + (1-p)N(\mu, \sigma^2)$, $X_{\bar{g}} \sim \bar{g}(.) := p\mathcal{N}(-\alpha, 1) + (1-p)\mathcal{N}(\alpha, 1)$, then, $S_g = S_{\bar{g}} + k_B \ln \sigma$ [22]. Similarly, for $p = \frac{1}{2}$, and $X_{\bar{f}} \sim \bar{f}(.) := \frac{1}{2}\mathcal{N}(-\alpha, 1) + \frac{1}{2}\mathcal{N}(\alpha, 1)$, we have $S_f = S_{\bar{f}} + k_B \ln \sigma$. Thus, $S_f - S_g = S_{\bar{f}} - S_{\bar{g}}$. Henceforth, we will consider $f_0(x) = Ce^{-\frac{(x+\alpha^2)}{2}}$ and $f_1(x) = Ce^{-\frac{(x-\alpha^2)}{2}}$, where $C = \frac{1}{\sqrt{2\pi}}$ and the symbol $x$ is overloaded and scaled with $\sigma$.

## 3. Upper and lower bounds on the change in entropy during erasure

For the derivation below it is assumed that $p \in [0.5, 1)$ and $\alpha > 1$. It follows (see Appendix) that,

$$\int_{-\infty}^{\infty} pf_0(x) \ln(pf_0(x) + (1-p)f_1(x))dx -$$
$$\int_{-\infty}^{\infty} pf_0(x) \ln(pf_0(x))dx$$
$$< C(2(1-p) + p \ln(\frac{e^{4\alpha^2}}{p}))e^{-\alpha^2/2}. \qquad (3)$$

Similarly, one can show that,

$$\int_{-\infty}^{\infty} (1-p)f_1(x)\ln(pf_0(x)+(1-p)f_1(x))dx$$
$$-\int_{-\infty}^{\infty} (1-p)f_1(x)\ln((1-p)f_1(x))dx$$
$$< C(2p + (1-p)\ln\left(\frac{e^{4\alpha^2}}{1-p}\right))e^{-\alpha^2/2}. \quad (4)$$

Let $K := \int_{-\infty}^{\infty} f_0(x)\ln(f_0(x))dx = \int_{-\infty}^{\infty} f_1(x)\ln(f_1(x))dx$. Using eq. (3) and eq. (4) with $p = \frac{1}{2}$ leads to the following lower bound on $S_f$,

$$S_f \geq k_B(-K + \ln 2 - C(2 + 4\alpha^2 + \ln 2)e^{-\alpha^2/2}). \quad (5)$$

From eq. (3) and (4), it also follows that,

$$-S_g \leq k_B(p\ln p + (1-p)\ln(1-p) + K$$
$$+ C(2 + p\ln p + (1-p)\ln(1-p) + 4\alpha^2)e^{-\alpha^2/2}). \quad (6)$$

In the Appendix, we also derive lower bounds for the difference,

$$\int_{-\infty}^{\infty} pf_0(x)\ln(pf_0(x)+(1-p)f_1(x))dx - \int_{-\infty}^{\infty} pf_0(x)\ln(pf_0(x))dx, \text{ and,}$$
$$\int_{-\infty}^{\infty} (1-p)f_1(x)\ln(pf_0(x)+(1-p)f_1(x))dx - \int_{-\infty}^{\infty} (1-p)f_1(x)\ln((1-p)f_1(x))dx.$$

Using eq. (16) and (17) with $p = 1/2$, we obtain the following upper bound on $S_f$,

$$S_f \leq k_B(-K + \ln 2 - \ln(1 + e^{-2\alpha^2})). \quad (7)$$

Furthermore, from eq. (16) and (17), we obtain the following lower bound on $-S_g$,

$$-S_g \geq k_B(K + p\ln p + (1-p)\ln(1-p) + p\ln(1 + \frac{1-p}{p}e^{-2\alpha^2}) + (1-p)\ln(1 + \frac{p}{1-p}e^{-2\alpha^2})). \quad (8)$$

It follows from eq. (5) and (7) that,

$$k_B(-K + \ln 2 - C(2 + 4\alpha^2 + \ln 2)e^{-\alpha^2/2})$$
$$\leq S_f \leq k_B(-K + \ln 2 - \ln(1 + e^{-2\alpha^2})). \quad (9)$$

Similarly, it follows from eq. (6) and (8) that,

$$k_B(K - H(p) + p\ln(1 + \frac{1-p}{p}e^{-2\alpha^2}) + (1-p)\ln(1 + \frac{p}{1-p}e^{-2\alpha^2}))$$
$$\leq -S_g \leq k_B(K - H(p) + C(2 - H(p) + 4\alpha^2)e^{-\alpha^2/2}), \quad (10)$$

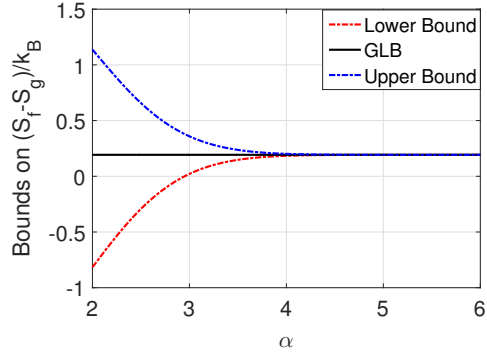

**Figure 6.** Lower and Upper bounds of $(S_f - S_g)/k_B$ and the GLB as a function of $\alpha$ for $p = 0.8$.

where, $H(p) = -p \ln p - (1-p) \ln (1-p)$. Using eq. (9) and (10), the difference between the initial and final entropy, $S_f - S_g$, satisfies the following bounds,

$$S_f - S_g \leq k_B(\ln 2 - H(p) - \ln(1 + e^{-2\alpha^2}) + C(2 - H(p) + 4\alpha^2)e^{-\alpha^2/2}), \text{ and,} \quad (11)$$

$$S_f - S_g \geq k_B(\ln 2 - H(p) - C(2 + 4\alpha^2 + \ln 2)e^{-\alpha^2/2} + p\ln(1 + \frac{1-p}{p}e^{-2\alpha^2}) +$$
$$(1-p)\ln(1 + \frac{p}{1-p}e^{-2\alpha^2})). \quad (12)$$

## 4. Relationship to the Generalized Landauer Bound

Note from eq. (11) and (12), $\lim_{\alpha \to \infty} S_f - S_g = k_B[\ln(2) + p\ln(p) + (1-p)\ln(1-p)]$(GLB); where the upper and lower bounds on the change in entropy, both converge to the GLB. The convergence to the limit is exponentially fast with regards to the overlap parameter $\alpha$ for the upper and lower bound. Thus, in the limiting case of $\alpha \to \infty$, it follows from the $2^{nd}$ Law of Thermodynamics that, $\frac{\langle Q_d \rangle}{T} \geq k_B[\ln(2) + p\ln(p) + (1-p)\ln(1-p)]$, implying that for a quasi static erasure with success proportion $p$, $\langle Q_d \rangle = k_B T[\ln(2) + p\ln(p) + (1-p)\ln(1-p)]$. In Figure 6, we present the above lower and upper bounds on $(S_f - S_g)/k_B$ as a function of the overlap parameter for $p = 0.8$. The exponential convergence to the GLB value for $p = 0.8$ is evident. Table 1 lists the difference between the upper (eq. (11)) and lower (eq. (12)) bounds of $(S_f - S_g)/k_B$ for various values of the overlap parameter, $\alpha$, and success proportion, $p$, where, $\alpha \approx 5$ is the seen as the transition point, beyond which, increasing $\alpha$(that is, the size of the memory bit) has insignificant gains from an energetics perspective. The recent experimental studies on verification of the Landauer's bound [8, 12, 9] had $\alpha \approx 20$, which indicates that there is scope to realize memory that is more closely 'packed'.

The case of $p = 1$ has to be considered separately. If $p = 1$, $g(x) = f_0(x)$ and $-S_g = k_B K$. It then follows from eq. (5) and (7) that,

$$S_f - S_g \geq k_B(\ln 2 - C(2 + 4\alpha^2 + \ln 2)e^{-\alpha^2/2}),$$
$$S_f - S_g \leq k_B(\ln 2 - \ln(1 + e^{-2\alpha^2})).$$

**Table 1.** Difference between the upper(eq. (11)) and lower(eq. (12)) bounds of $(S_f - S_g)/k_B$ as a function of $\alpha$ and $p$.

| $\alpha$ \ $p$ | 0.6 | 0.7 | 0.8 |
|---|---|---|---|
| 1.5 | 2.829 | 2.838 | 2.852 |
| 2.0 | 1.944 | 1.947 | 1.953 |
| 2.5 | 0.946 | 0.948 | 0.949 |
| 3.0 | 0.336 | 0.337 | 0.338 |
| 5.0 | $3.033e^{-4}$ | $3.034e^{-4}$ | $3.035e^{-4}$ |

In the case $p = 1$ as well, it follows that $\lim_{\alpha \to \infty} S_f - S_g = k_B \ln(2)$, which is the Landauer's bound.

## 5. Extensions to Asymmetric 1 bit Memory

In this section, we extend the GLB to asymmetric single bit memory, where the two states are non identical wells. An asymmetric one bit memory and the associated minimum heat dissipation for its erasure is discussed in [19, 11]. In particular, [11] presents an experimental study of the minimum heat dissipation for perfect erasures of a bit of asymmetric memory, with one well being wider than the other, and considers the two cases of resetting the bit into the wider and narrower well. Motivated from the discussion in [11], we extend the GLB to the case of non-identical wells, where, one well is wider than the other.

We assume that initially, the particle has equal probability to be in either well and the initial probability distribution of the particle is given by $f(x)$ as described earlier (see eq. (1)). Similarly, for erasures with success proportion $p$, the final probability distribution of the particle is given as $g(x)$ (see eq. (2)).

### 5.1. Erasing into low entropy well

Consider $f_0(x) = Ce^{-\frac{(x+\mu)^2}{2\sigma^2}}$ and $f_1(x) = \frac{C}{\beta}e^{-\frac{(x-\mu)^2}{2(\beta\sigma)^2}}$ with $\beta > 1$. The particle has higher entropy in the state 1 as compared to state 0. In this case, for 'reset to zero' with success proportion $p$, once can show that $\lim_{\alpha \to \infty} S_f - S_g = k_B[p \ln p + (1-p)\ln(1-p) + (\frac{1}{2} - (1-p))\ln\beta + \ln 2]$. The rate of convergence is exponential with respect to the parameter $\alpha$ and depends inversely on the parameter $\beta$. Thus, in the limiting case of the two wells being sufficiently far apart and a quasi-static erasure process with success proportion $p$, $\langle Q_d \rangle = k_B T[p \ln p + (1-p)\ln(1-p) + (\frac{1}{2} - (1-p))\ln\beta + \ln 2]$. Here, the bit was erased into the well with lower entropy, hence, the limiting value of the change in entropy is higher than the identical wells case. In Figure 7 the heat dissipation associated with erasing into the low entropy well is shown for various values of the asymmetry parameter $\beta$ and is compared with the Generalized Landauer Bound. It is seen that, as the asymmetry parameter is increased, the associated heat dissipation also increases for various success proportions. The case of erasing into the high entropy well is much more interesting and is presented next.

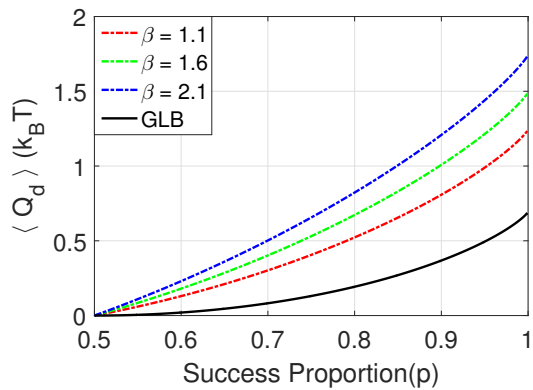

**Figure 7.** Minimum heat dissipation associated with erasing into low entropy well in a quasi static manner.

### 5.2. Erasing into high entropy well

Consider erasing into the higher entropy well, that is, $f_1(x) = Ce^{-\frac{(x-\mu)^2}{2\sigma^2}}$ and $f_0(x) = \frac{C}{\beta}e^{-\frac{(x+\mu)^2}{2(\beta\sigma)^2}}$ with $\beta > 1$. Here, the particle has higher entropy in the state 0 as compared to state 1. In this case, for 'reset to zero' with success proportion $p$, once can show that $\lim_{\alpha \to \infty} S_f - S_g = k_B[p \ln p + (1-p)\ln(1-p) + ((1-p) - \frac{1}{2})\ln \beta + \ln 2]$. The rate of convergence is exponential with respect to the parameter $\alpha$ and depends inversely on the parameter $\beta$. Thus, in the limiting case of the two wells being sufficiently far apart, $\langle Q_d \rangle = k_B T[p \ln p + (1-p)\ln(1-p) - (\frac{1}{2} - (1-p))\ln \beta + \ln 2]$. Here, the bit was erased into the well with higher entropy, hence, the limiting value of the change in entropy is lower than the identical wells case. In Figure 8 the heat dissipation associated with erasing into the low entropy well is shown for various values of the asymmetry parameter $\beta$ and is compared with the Generalized Landauer Bound. It is seen that, as the asymmetry parameter is increased, the associated heat dissipation also increases for various success proportions. In this case, it is seen that erasure can be achieved without any associated heat dissipation. Hence, asymmetry can be utilized toward realizing logically irreversible computations with zero dissipation.

### 6. Conclusion

We quantify the dependence of the decrease in entropy associated with erasure of a bit of information on the amount of overlap between the equilibrium distributions of the two states of a one bit memory. It is seen that overlap can lead to considerably lower heat dissipation as compared to the GLB in a quasi static erasure process. In particular, we present quantitative bounds on the change in entropy associated with partial information erasure, which exponentially converge to the GLB when the physical separation between the two states is large. A conclusion is reached that $\alpha \approx 5$ represents a threshold for energetics of computation associated with a single bit memory, and memory bit designs with $\alpha > 5$ have insignificant gains compared to $\alpha \approx 5$ case from an energetics perspective. This could possibly be a memory bit design guideline, which might lead to improvements in memory density. Furthermore, we discussed the effect of asymmetry between the two wells of a single bit memory on the associated minimum heat dissipation for erasure processes. It is seen that, the decrease in entropy associated with erasing into the state with lower volume, results in minimum heat dissipation for a quasi static erasure process being higher than the GLB. However, erasing into the state with higher volume, results in the associated minimum heat dissipation for a quasi static erasure process being not only lower than the GLB but also with no associated dissipation. Thus, physical modifications in a single bit memory either due to overlap between the two states or due to asymmetry between the two states, along with uncertainty in the outcome of the erasure process, are some mechanisms to realize erasures with heat dissipation lower than the Landauer bound.

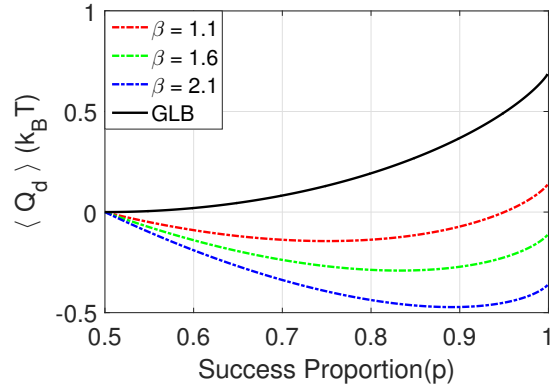

**Figure 8.** Minimum heat dissipation associated with erasing into high entropy well in a quasi static manner.

## 7. Acknowledgments

We would like to thank Prof. Arnab Sen, Department of Mathematics, University of Minnesota for initial discussion about the problem. The authors acknowledge the support of the National Science Foundation for funding the research under Grant No. CMMI-1462862.

## 9. Appendix: Derivation of eq. (3)

*9.1. Upper Bounds*

First, we derive an upper bound for the difference,

$$\int_{-\infty}^{\infty} pf_0(x) \ln(pf_0(x) + (1-p)f_1(x))dx -$$
$$\int_{-\infty}^{\infty} pf_0(x) \ln(pf_0(x))dx,$$

which is simplified as,

$$p\int_{-\infty}^{\infty} f_0(x) \ln(1 + \frac{(1-p)f_1(x)}{pf_0(x)})dx$$
$$= Cp\int_{-\infty}^{\infty} e^{-t^2/2} \ln(1 + \frac{(1-p)}{p}e^{2(t-\alpha)\alpha})dt, \, (t := x + \alpha),$$
$$= Cp\int_{(-\infty,\alpha)\cup[3\alpha,\infty)} e^{-t^2/2} \ln(1 + \frac{(1-p)}{p}e^{2(t-\alpha)\alpha})dt$$
$$+ Cp\int_{\alpha}^{3\alpha} e^{-t^2/2} \ln(1 + \frac{(1-p)}{p}e^{2(t-\alpha)\alpha})dt. \tag{13}$$

Using the logarithmic inequality $\ln(1+x) < x$ for $x > 0$, the first integral in eq. (13) satisfies,

$$Cp\int_{(-\infty,\alpha)\cup[3\alpha,\infty)} e^{-t^2/2} \ln(1 + \frac{(1-p)}{p}e^{2(t-\alpha)\alpha})dt$$
$$< C(1-p)\int_{(-\infty,\alpha)\cup[3\alpha,\infty)} e^{-(t-2\alpha)^2/2}dt,$$
$$= 2C(1-p)\int_{\alpha}^{\infty} e^{-u^2/2}du, \, (u := t - 2\alpha),$$
$$< 2C(1-p)\int_{\alpha}^{\infty} ue^{-u^2/2}du, \, (\because u \geq \alpha > 1),$$
$$= 2C(1-p)e^{-\alpha^2/2}. \tag{14}$$

Moreover, $0 \leq 2(t-\alpha)\alpha \leq 4\alpha^2$ for $t \in [\alpha, 3\alpha)$ and $e^{4\alpha^2} > 1$, using which the last integral in eq. (13) satisfies,

$$Cp\int_{\alpha}^{3\alpha} e^{-t^2/2} \ln(1 + \frac{(1-p)}{p}e^{2(t-\alpha)\alpha})dt,$$
$$\leq Cp\int_{\alpha}^{3\alpha} e^{-t^2/2} \ln(\frac{e^{4\alpha^2}}{p})dt,$$
$$\leq Cp\ln(\frac{e^{4\alpha^2}}{p})\int_{\alpha}^{\infty} e^{-t^2/2}dt,$$
$$\leq Cp\ln(\frac{e^{4\alpha^2}}{p})e^{-\alpha^2/2}. \tag{15}$$

The inequality in eq. (3) then follows by combining eq. (14) and (15) with (13).

*9.2. Lower Bounds*

We now derive a lower bound for the difference,

$$\int_{-\infty}^{\infty} p f_0(x) \ln(p f_0(x) + (1-p) f_1(x)) dx - \int_{-\infty}^{\infty} p f_0(x) \ln(p f_0(x)) dx$$
$$= \int_{-\infty}^{\infty} p \frac{e^{-x^2/2}}{\sqrt{2\pi}} \ln(1 + \frac{(1-p)}{p} e^{2(x-\alpha)\alpha}) dx.$$

Notice that by direct computation of second derivatives, it follows that the function $\varphi(t) = p \ln(1 + \frac{1-p}{p} e^{2(t-\alpha)\alpha})$ is a convex function. Thus, we can express,

$$\int_{-\infty}^{\infty} \frac{e^{-x^2/2}}{\sqrt{2\pi}} p \ln(1 + \frac{(1-p)}{p} e^{2(x-\alpha)\alpha}) dx = \mathbb{E}(\varphi(Z)),$$

where, $Z$ is a standard Gaussian random variable. Applying Jensen's inequality [22], we have,

$$\mathbb{E}(\varphi(Z)) \geq \varphi(\mathbb{E}(Z)),$$
$$= p \ln(1 + \frac{1-p}{p} e^{-2\alpha^2}). \tag{16}$$

Similarly, one can show that,

$$\int_{-\infty}^{\infty} (1-p) f_1(x) \ln(p f_0(x) + (1-p) f_1(x)) dx - \int_{-\infty}^{\infty} (1-p) f_1(x) \ln((1-p) f_1(x)) dx$$
$$= \int_{-\infty}^{\infty} \frac{e^{-x^2/2}}{\sqrt{2\pi}} (1-p) \ln(1 + \frac{p}{1-p} e^{-2(x+\alpha)\alpha}) dx$$
$$= \mathbb{E}(\Psi(Z)),$$

where, $\Psi(t) = (1-p) \ln(1 + \frac{p}{1-p} e^{-2(t+\alpha)\alpha})$ is a convex function. Using Jensen's inequality,

$$\mathbb{E}(\Psi(Z)) \geq \Psi(\mathbb{E}(Z))$$
$$= (1-p) \ln(1 + \frac{p}{1-p} e^{-2\alpha^2}). \tag{17}$$